\documentclass[a4paper,aps,prl,showpacs,twocolumn,floatfix,superscriptaddress]{revtex4}
\usepackage{graphicx} % Include figure files
\usepackage{amsmath}
\usepackage{bm}
\usepackage{dcolumn}% Align table columns on decimal point
\begin{document}

\title{Trapping of Neutral Mercury Atoms and Prospects for Optical Lattice Clocks}

\author{H. Hachisu}
\affiliation{Department of Applied Physics, Graduate School of Engineering, The University of Tokyo,
Bunkyo-ku, Tokyo 113-8656, Japan}
\affiliation{CREST, Japan Science and Technology Agency, 4-1-8 Honcho Kawaguchi, Saitama, Japan}

\author{K. Miyagishi}
\affiliation{Department of Applied Physics, Graduate School of Engineering, The University of Tokyo,
Bunkyo-ku, Tokyo 113-8656, Japan}

\author{S. G. Porsev}
\affiliation{Petersburg Nuclear Physics Institute, Gatchina,
Leningrad district, 188300, Russia}
 \affiliation{Physics Department, University of
Nevada, Reno, Nevada 89557, USA}

\author{A. Derevianko }
 \affiliation{Physics Department, University of
Nevada, Reno, Nevada 89557, USA}
 \affiliation{Laboratoire Aim\'{e} Cotton, B\^{a}t. 505, Campus d'Orsay, 91405 ORSAY Cedex France}

\author{V. D. Ovsiannikov}
 \affiliation{Physics Department, Voronezh State University, Universitetskaya pl. 1, 394006, Voronezh, Russia}

\author{V. G. Pal'chikov}
 \affiliation{Institute of Metrology for Time and Space at National Research Institute for Physical-Technical and Radiotechnical Measurements, Mendeleevo, Moscow Region, 141579, Russia
}

\author{M. Takamoto}
\affiliation{Department of Applied Physics, Graduate School of Engineering, The University of Tokyo,
Bunkyo-ku, Tokyo 113-8656, Japan}
\affiliation{CREST, Japan Science and Technology Agency, 4-1-8 Honcho Kawaguchi, Saitama, Japan}

\author{H. Katori}
\affiliation{Department of Applied Physics, Graduate School of Engineering, The University of Tokyo,
Bunkyo-ku, Tokyo 113-8656, Japan}
\affiliation{CREST, Japan Science and Technology Agency, 4-1-8 Honcho Kawaguchi, Saitama, Japan}

\date{\today}
\begin{abstract}
We report a vapor-cell magneto-optical trapping of Hg isotopes on the ${}^1S_0-{}^3P_1$ intercombination transition.
Six abundant isotopes, including four bosons and two fermions, were trapped. 
Hg is the heaviest non-radioactive atom trapped so far, which enables sensitive atomic searches for ``new physics'' beyond the standard model.
We propose an accurate optical lattice clock based on Hg and evaluate its systematic accuracy to be better than $10^{-18}$.
Highly accurate and stable Hg-based clocks will provide a new avenue for the research of optical lattice clocks and the time variation of the fine-structure constant.
\end{abstract}
\maketitle

%32.10.Dk Electric and magnetic moments, polarizability
%06.30.Ft Time and frequency
%31.25.-v Electron correlation calculations for atoms and molecules
Precision tests of fundamental symmetries with atoms and molecules play an important role in
probing new physics beyond the standard model of elementary particles.
The ``new physics'' sensitivity
of such experiments depends steeply on the nuclear charge $Z$ of the atom. 
In particular,
 parity-violating amplitudes change as $Z^3$~\cite{BouBou74},
 atomic CP-violating (T-, P-odd) permanent electric dipole moments (EDM), as $Z^3$~\cite{San65},
 and CP-violating polarizabilities, as $Z^5$~\cite{RavKozDer05}. 
As a result, the most accurate limit to date on the atomic EDM has been set with Hg atoms ($Z=80$) and on the electron EDM with Tl ($Z=81$) atoms \cite{RomGriJac01,RegComSch02}.
In a quest for more sensitive probes, experiments with heavier atoms, Fr ($Z=87$) and Ra ($Z=88$), have been proposed~\cite{GomOroSpr06,GueSciAhm07etal}. 
However, all the elements with $Z>83$ are radioactive and consequently experience associated drawbacks.

In this Letter, we report the vapor-cell magneto-optical trapping (MOT) of Hg isotopes and discuss its application to accurate optical lattice clocks \cite{Kat03}. 
Located at the very edge of the stability valley, Hg is, so far, the heaviest non-radioactive atom that has been laser-cooled and trapped.
It is anticipated that these ultracold heavy atoms will enable experimental searches of high-$Z$ atom physics with unprecedented sensitivity due to reduced motional effects, longer interrogation time, and higher optical thickness. Moreover, as demonstrated for Yb isotopes \cite{Takahashi07}, a remarkable variety of trapped Hg isotopes (4 bosonic and 2 fermionic ones) may allow attaining Bose or Fermi-degeneracy, which would further improve the measurements' sensitivity.

We systematically evaluate various sources of uncertainty for the Hg-based optical lattice clock and argue that an accuracy of  better than $10^{-18}$ is attainable, which is an order of magnitude of improvement over Sr \cite{Kat03} or Yb-based \cite{PorDer06BBR} clocks because of reduced susceptibility to the blackbody radiation field, which sets a major limitation on the accuracy of atomic clocks \cite{PorDer06BBR,NIST07}.
This projected accuracy is competitive with that of the best ion clock with Al$^+$~\cite{NIST07}.
In particular, we take advantage of the high nuclear charge $Z$ of Hg to explore the variation of the fine-structure constant $\alpha$ that generically controls the strength of relativistic effects in the atom as $(\alpha Z)^2$.
While astrophysical observations suggest that $\alpha$ may drift at the rate of $\dot{\alpha}/\alpha\approx 10^{-16}\,\mathrm{yr}^{-1}$ \cite{MurFlaWeb04}, recent laboratory tests based on the frequency ratio of accurate ion clocks and Cs clocks set the upper bound of $|\dot{\alpha}/\alpha|<1.3 \times 10^{-16}\,\mathrm{yr}^{-1}$ \cite{Fortier07} limited by Cs clock's inaccuracy of $1\times 10^{-15}$. 
Significant enhancement in measurement precision will be expected for direct comparison of optical clocks with  frequency comb technique \cite{ma2004ofs}: the frequency change $\delta \nu$ of Hg-based clocks operated at $\nu_0$ can be measured in reference to Sr \cite{Kat05} or Hg$^+$ \cite{Fortier07} clocks, where Hg, Sr, and Hg$^+$ have sensitivity $(\delta \nu/\nu_0)/(\delta\alpha/\alpha)$ of $0.81$, $0.06$, and $-3.19$, respectively, to the fractional change of $\delta\alpha/\alpha$ \cite{MurFlaWeb04,Ang04}.

\begin{figure}[h]
\begin{center}
\includegraphics[width=0.9\linewidth]{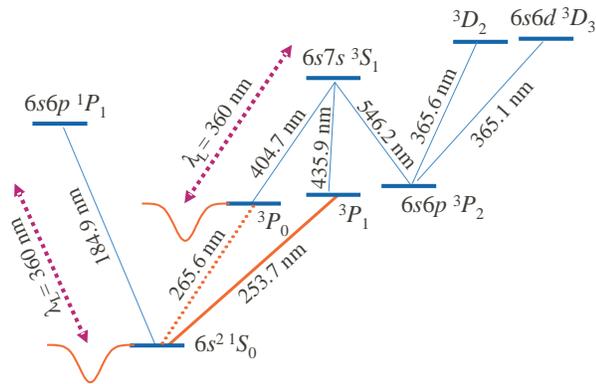}
\caption{Energy-level diagram for Hg. The $^1\!S_0-{}^3\!P_1$ transition at 253.7~nm is used for magneto-optical trapping of the atoms. The $^1S_0-{}^3P_0$ transition at 265.6~nm serves as a clock oscillator.
The ``magic wavelength,'' $\lambda_L=360~{\rm nm}$, is the far-off resonance for the relevant transitions originating from the $^1S_0$ and $^3P_0$ levels.
}
\label{fig1}
\end{center}
\end{figure}

The laser cooling and clockwork with neutral Hg exploit its divalent electronic structure and parallel techniques developed for Sr \cite{Kat99,Kat03}.
The energy levels of Hg are shown in Fig.~1.
The spin-forbidden $^1\!S_0-{}^3\!P_1$ transition at 253.7~nm with the natural linewidth of $\gamma/2\pi=1.3$~MHz \cite{HgLine} provides an adequate cooling transition down to the Doppler temperature of $31~\mu$K.
The $^1\!S_0-{}^3\!P_0$ transition at 265.6~nm, which is weakly allowed for odd isotopes with a nuclear spin ($I\neq0$) or even isotopes $(I=0)$ dressed by external fields \cite{Tai06,Pal07}, will be exploited as a clock transition. 
The ``magic wavelength,'' where the light-shift perturbations of the clock levels are canceled out \cite{Kat99JPSJ,Kat03}, is calculated to be $\lambda_L=360~{\rm nm}$, as discussed later.

Trapping neutral Hg is experimentally challenging due to limited laser power available at deep-UV cooling transition wavelength.
To overcome this difficulty and take advantage of its heavy mass and the large radiation pressure at the shorter wavelengths, we used a vapor-cell MOT~\cite{LinWie92} to capture the atoms from the low-velocity tail of the Maxwell-Boltzmann distribution. Our apparatus is shown in Fig.~2.
A 253.7~nm light was generated by frequency quadrupling a laser diode system operated at 1014.9~nm.
A master laser was an external-cavity laser diode locked to a reference cavity to reduce its linewidth to less than 100~kHz.
This master laser was amplified by a tapered amplifier up to 1 W and injected into a bow-tie cavity with a PPLN crystal to generate 240~mW at 507.4~nm. This was further frequency-doubled by a BBO crystal to obtain roughly 10~mW at 253.7~nm.

This UV light was fed into a vacuum chamber with its base pressure of $3\times10^{-10}$~Torr to form a vapor-cell MOT with a folded beam configuration with a single stroke.
The $1/e^2$ diameter of the trapping beam was 20~mm with a power density of $0.3 I_0$, where $I_0=10\,{\rm mW/cm}^2$ is the saturation intensity of the $^1S_0-{}^3P_1$ transition.
After introducing Hg, the vacuum pressure was increased to $8\times 10^{-9}$~Torr that indicated the Hg-atom number density of $7\times 10^7~{\rm cm}^{-3}$.
The typical magnetic field gradient of a quadrupole magnetic field was 10~G/cm along its axis or the Zeeman shift gradient of 21~MHz/cm for the $^3P_1$ state.

\begin{figure}[h]
\begin{center}
\includegraphics[width=0.9\linewidth]{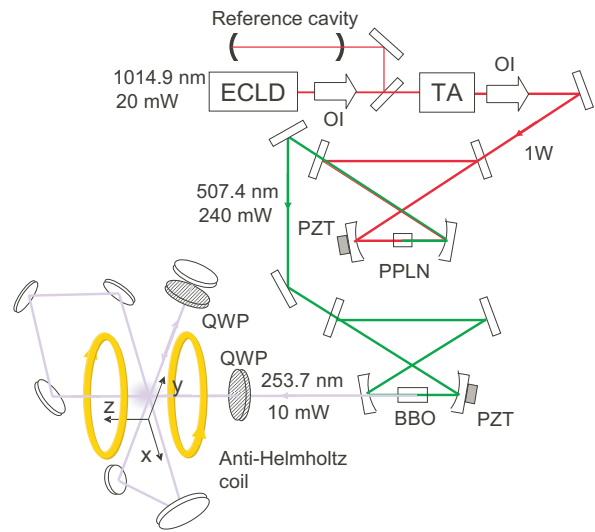}
\caption{Experimental setup. A 253.7~nm UV light was generated by frequency quadrupling a diode laser operated at 1014.9~nm with nearly 1~W. A 10~mW UV light was fed into a vacuum chamber to form a vapor-cell MOT with a folded beam configuration. ECLD: External Cavity Laser Diode, OI: Optical Isolator, TA: Tapered Amplifier, PZT: Piezo-electric Transducer, QWP: Quarter Wave Plate.
}
\label{fig2}
\end{center}
\end{figure}

Hg has 6 stable isotopes with abundances of more than $6\%$; 4 bosonic  ($^{198}$Hg, $^{200}$Hg, $^{202}$Hg, and $^{204}$Hg) isotopes with nuclear spin $I=0$ and 2 fermionic $^{199}$Hg and $^{201}$Hg isotopes with $I=1/2$ and $3/2$, respectively.
The spectra of trapped Hg isotopes are shown in Fig.~3, where the UV frequency was swept toward higher frequencies with 0.5~MHz/s. The typical atom filling time was measured to be $\sim 1$~s for $^{202}$Hg. 
The trap spectra are about 10~MHz or $8\gamma$ wide, whose asymmetry are explained in terms of atom loading dynamics into the MOT \cite{LinWie92,Raab87}.
While atom capturing efficiency is higher in the red wing, the trapped atoms scatter more photons as the laser detuning $\Delta$ becomes smaller. The atoms suddenly boil up from the MOT when they cross the Doppler cooling condition of $\Delta=-\gamma/2$ over to their resonance, corresponding to the sharp cutoff in the blue wing.

In Fig.~3, the number of trapped isotopes was nearly proportional to their natural abundance, indicating that atoms can be further accumulated until two-body light-assisted collisions limit their density.
For odd isotopes (with a hyperfine structure), stable MOTs were observed for the $^1S_0(F=1/2)\rightarrow{}^3P_1(F=3/2)$ transition for $^{199}$Hg and $^1S_0(F=3/2)\rightarrow{}^3P_1(F=5/2)$ for $^{201}$Hg, as expected \cite{Raab87}.
%A weak MOT was also observed for the $F=1/2\rightarrow F=1/2$ transition in $^{199}$Hg, which is due to random optical pumping of the $^1S_0(F=1/2)$ %ground state by trapping lasers irradiated from transverse directions.
From the fluorescence of the MOT, the number of trapped atoms was roughly $10^6$ for $^{202}$Hg with a $1/e^2$ cloud radius of about 0.2~mm.
The atomic temperature was measured to be less than $50~\mu$K by the time-of-flight technique, which was consistent with its Doppler temperature.
Even lower temperatures are expected for odd isotopes because of the Sisyphus cooling mechanism \cite{PGC}.
Since atom loading efficiency into the vapor-cell MOT strongly depends on trapping laser diameter \cite{LinWie92}, substantial increase in the number of trapped atoms may be expected by applying brighter light sources \cite{ELS07}.

\begin{figure}[h]
\begin{center}
\includegraphics[width=0.9\linewidth]{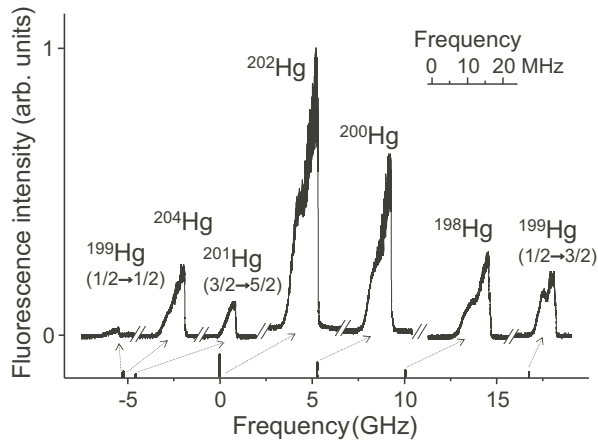}
\caption{Fluorescence spectrum of magneto-optically
trapped Hg isotopes as a function of the laser frequency at 253.7~nm.
The bars at the bottom indicate the positions of the resonances of individual isotopes. Detailed profiles are drawn 
using the magnified frequency scale shown at the top.
}
\label{fig3}
\end{center}
\end{figure}

In the following, we discuss an ultrastable optical lattice clock operating on the $6s^2 \,^1\!S_0-6s6p \,^3\!P_0$ transition (see Fig.~1) of Hg.
Starting with the magneto-optically trapped atomic samples, as described above, a possible experimental scenario is as follows: a 100-$\mu$K-deep 1D (one-dimensional) optical lattice is prepared inside a power build-up cavity with a peak intensity of the standing wave of $I_L=1~{\rm MW/cm}^2$, where the light-shift $\delta \nu_L$ is given by $\delta \nu_L/I_L\approx 2~\mathrm{kHz/(kW/cm^2)}$.
The trapped atoms are then sideband-cooled to the vibrational ground state of the fast axis on the $^1S_0-{}^3P_0$ clock transition with an appropriate relaxation process via the $^3S_1$ state (see Fig.~1). 
The optical lattice clock can be operated after adiabatically reducing the lattice intensity down to $I_L={\rm 50~kW/cm}^2$, which corresponds to the potential depth of $\sim 10 E_R$, in order to minimize the hyperpolarizability effects of the lattice lasers. 
Here, $E_R=(h/\lambda_L)^2/2m$ is the photon recoil energy of Hg atoms scaled by the magic wavelength $\lambda_L$.
In Ref.~\cite{Lem05}, it is pointed out that optical lattices as shallow as $10 E_R$ could well meet the $10^{-18}$ accuracy goal by suppressing the tunneling of atoms with the gravitational acceleration when the 1D lattice is vertically oriented.

We evaluate the accuracy of the Hg optical lattice clock 
by carrying out relativistic calculations of the relevant atomic properties.
Hg has 80 electrons, and the correlations
play an essential role in characterizing the atomic properties.
We employ the method of configuration-interaction (CI) coupled
with the many-body perturbation theory (MBPT) \cite{DzuFlaKoz96}. 
%The calculations are carried out in the $V^{N-1}$ basis.
The basis set was constructed in the V$^{(N-1)}$ approximation.
In CI+MBPT, the
positions of the low-lying energy
levels are reproduced within a few 0.1\%. Hyperfine-structure constants and dipole matrix elements have a theoretical accuracy better than 10\%.
The most relevant test for the prediction of the ``magic wavelength'' is the difference in the static polarizabilities of the $6s6p\,^3\!P_1$ and the ground states, measured in Ref.~\cite{HarRom00}
to be 26.66(47) a.u.
Our computed value of 26.95 a.u. is in agreement with this experiment.

We start by determining the natural width of the clock transition.
Although a single-photon
$6s^2 \,^1\!S_0-6s6p \,^3\!P_0$
transition is forbidden by the conservation of the angular momentum, hyperfine coupling with the nuclear angular momentum $I\neq0$  opens a weak decay channel~\cite{PorDer04}.
Our calculations of the hyperfine quenching result in the rate coefficients
$A (^{199}\mathrm{Hg} ) =  1.3 \times 10^{-2}\, \mathrm{s}^{-1}$ and
$A (^{201}\mathrm{Hg} ) =  8.8 \times 10^{-3}\, \mathrm{s}^{-1}$. 
In even isotopes ($I$=0), the natural linewidth would be introduced only due to highly suppressed E1-M1 two-photon decays.

In an optical lattice, the shift of atomic levels is proportional to the a.c.~polarizability $\alpha( \lambda)$,
and we compute it for both clock levels. They cross at
the magic wavelength of $\lambda_L \approx 360 \, \mathrm{nm}$, which
agrees with the semi-empirical estimates of 358 nm~\cite{Kat05}
and of 342 nm~\cite{OvsPalTak06}.
Compared to the E1 contribution,
the higher-rank multipolar a.c.~polarizabilities are suppressed,
and they merely shift $\lambda_L$ by a small fraction.

%The non-scalar nature of  also affects the a.c.~polarizability. 
In addition to the dominant scalar contribution, the fermionic isotopes ($I \neq 0$) acquire 
vector $\alpha^{(a)}(\lambda_L)$ (and tensor for $I>1/2$) components of the a.c.~polarizability.
The non-scalar contributions are
suppressed by the ratio of the hyperfine interaction to atomic energy.
The $^3P_0$ level is affected the most due to the nearby levels of the  $^3P_J$ fine-structure manifold. We find
$\alpha^{(a)}(\lambda_L) = 0.040$ a.u.~for $^{199}\mathrm{Hg}$ and $\alpha^{(a)}(\lambda_L) = -0.045$ a.u.~for $^{201}\mathrm{Hg}$.
Such vector shifts can be eliminated using an averaging procedure~\cite{JPSJ06}:
Since the vector light shift is proportional to the $m_F$ of the Zeeman sublevels, alternating measurements of $f_{\pm}$ of the $^1S_0(F,\pm m_F)- {}^3P_0(F,\pm m_F)$ transition frequencies and taking the average $f_0=(f_{+}+f_{-})/2$ can significantly reduce the fractional contribution $\rho$ of the vector light-shift down to $\delta \nu_v\approx \nu_x \rho^2$, where $\nu_x\approx 50\,\mathrm{kHz}$ is the lattice trapping frequency for the $10 E_R$-deep lattice.
By applying nearly linearly polarized light with its ellipticity of less than $10^{-2}$, $\delta \nu_v\sim 10^{-5}$~Hz makes a negligible contribution to our goal accuracy.

We also made a semi-empirical estimate of the
second-order (in the lattice intensity $I_L$) corrections to the a.c.~Stark shift of the clock frequency
 $\delta \nu_h= \Re(\beta)\, I_L^2, $
 where $\beta=(\gamma_{^1S_0}-\gamma_{^3P_0})/64$ is the difference between the
 clock-level hyperpolarizabilities and $\Re(\beta)$ denotes its real part.
The calculations in the single-electron approximation, as carried out in Sr~\cite{Kat03}, infer that $\gamma_{^3P_0}$ is about $10^2$ times larger than $\gamma_{^1S_0}$ at $\lambda_L=360$~nm.
It becomes smaller with decreasing $\lambda_L$ and crosses
zero near 347 nm for a linear polarization. Numerically, $\beta
=-(103+5.5i)~{\rm mHz/(MW/cm}^2)^2$ for $\lambda_L=360$ nm, and
$\beta =-(21.7+1.5i)~{\rm mHz/(MW/cm}^2)^2$ for $\lambda_L=351$
nm. The imaginary part of $\beta$ determines the two-photon
ionization rate $P_\mathrm{2ph}=-4\pi\Im(\beta) \, I_L^2$.
For $I_L$=~50~kW/cm$^2$ and the linear polarization, the induced shift $\delta \nu_h$ is below
0.3 mHz, and the two-photon ionization rate does not exceed
$2 \times 10^{-4}~{\rm s}^{-1}$, which is smaller than the off-resonant (single-) photon scattering rate of
 $5\times10^{-2}\,{\rm s}^{-1}$ of the lattice lasers. For the circular polarization, both the real and the
imaginary parts of $\beta$ are several times larger,
without any node in the indicated region of wavelengths.
These second-order corrections are sensitive to both correlations and the magic wavelength of $\lambda_L$; therefore, they ultimately need to be investigated experimentally.

The sensitivity to magnetic fields $B$ is determined by the difference in the Land\'{e} g-factors of the ground and excited states.
We find 
$\delta g (^{199}\mathrm{Hg} ) = -4.7 \times 10^{-4}$ and
$\delta g (^{201}\mathrm{Hg} ) =  1.8 \times 10^{-4}$ with the CI+MBPT calculations.
While the linear Zeeman shift can be eliminated by the same averaging protocol
as described previously \cite{JPSJ06}, the simultaneously obtained Zeeman shift $\Delta f=f_{+}-f_{-}$ with $\delta g$ given above provides a real-time probe of the applied magnetic field to correct the second-order Zeeman effects of
 $\delta \nu_Z^{(2)}=-\beta_Z B^2$ with $\beta_Z=24.4 \mathrm{mHz/G}^2$.

Finally, at room temperature, the accuracy of the clock is affected by 
blackbody radiation (BBR). The induced shift of the clock transition frequency,
$\delta \omega_\mathrm{BBR} \approx -\frac{2}{15} (\alpha \pi)^3 T^4 \times
( \alpha_{^3\!P_0}-  \alpha_{^1\!S_0}) $, depends on the difference in
the static polarizabilities of the clock states. Our CI+MBPT calculations result in
$ \alpha_{^3\!P_0} =54.6$ a.u.~and $\alpha_{^1\!S_0}=33.6$ a.u., leading to
the BBR shift of $\delta \nu_\mathrm{BBR} \approx -0.181\,\mathrm{Hz}$ at $T=300\, \mathrm{K}$.
It is noteworthy that the fractional
correction $\delta \nu_{\rm BBR}/\nu_0 \approx -1.60 \times 10^{-16}$ is an order
of magnitude smaller than the Sr and Yb based clocks~\cite{PorDer06BBR}.
Assuming a temperature uncertainty of 0.1~K, the fractional uncertainty due to the BBR is $2\times10^{-19}$.

In conclusion, we have demonstrated a magneto-optical trapping of six isotopes of the Hg atom, the heaviest non-radioactive atom trapped so far.
We have shown that Hg is a promising candidate for highly accurate optical lattice clocks with an estimated inaccuracy of less than $10^{-18}$, which is mainly limited by the hyperpolarizability effects, higher-rank multipolar polarizabilities, and BBR shifts.
% and improves the uncertainties of Sr and Yb-based clocks by an order of magnitude.
Although, current performances of optical lattice clocks have been investigated at the $10^{-15}$ level \cite{JPSJ06,Boyd07,Tar06} limited by the uncertainty of Cs reference clocks, a highly accurate and stable Hg-based clock will provide a new avenue for improved systematic studies of optical lattice clocks, such as the precise determination of the ``magic wavelength'' and other properties beyond the $1\times 10^{-15}$ uncertainty level.
Furthermore, because of its high $Z$, Hg may allow a sensitive search of the temporal variation of $\alpha$ combined with other optical clocks, e.g., Sr-based clocks as a possible anchor \cite{Kat05}.

We would like to thank Y. Tanino and S. I. Marmo.
This work was supported in part by SCOPE, NSF, RFBR (Grant Nos.~07-02-00210-a and 07-02-00278), CRDF, and MinES.

%\bibliography{all,HgclockAddOn}

\end{document}